\documentclass[reprint, amsmath, amssymb, aps]{revtex4-2}

\usepackage{dcolumn}% Align table columns on decimal point
\usepackage{bm}

\usepackage[T1]{fontenc}
\usepackage{indentfirst}
\usepackage{xcolor}
\usepackage{graphicx}
\usepackage{amssymb}
\usepackage{amsmath}
\usepackage{latexsym}
\usepackage{mathtools}
\usepackage{bbold}
\usepackage{accents}
\usepackage{hyperref} % for web links
\usepackage[utf8]{inputenc} 
\usepackage{multirow}
\usepackage{palatino} 
\usepackage{lmodern}
\usepackage{soul}
\usepackage{array}
    \newcolumntype{P}[1]{>{\centering\arraybackslash}p{#1}}

\usepackage{adjustbox}

\newcommand\Nc{\mathbf{\hat{N}_{c}}}
\newcommand\hatsc{\mathbf{\hat{s}_{c}}}
\newcommand\hatD{\mathbf{\hat{\Delta}}}
\newcommand\Ctilde{\mathbf{\Tilde{C}}}

\DeclareMathOperator*{\argmax}{arg\,max}

\begin{document}

\title{Forecasting the performance of the Minimally Informed foreground cleaning method for CMB polarization observations}

\author{Clément Leloup$^{1,2}$}
\email{clement.leloup@ipmu.jp}
\author{Magdy Morshed$^{3,4,5}$}
\author{Arianna Rizzieri$^{6}$}
\affiliation{$^{1}$ Kavli Institute for the Physics and Mathematics of the Universe (WPI), The University of Tokyo \\
 Institutes for Advanced Study,The University of Tokyo, Kashiwa, Chiba 277-8583, Japan}
\affiliation{$^{2}$ Center for Data-Driven Discovery, Kavli IPMU (WPI), UTIAS, The University of Tokyo, Kashiwa, Chiba 277-8583, Japan}
\affiliation{$^{3}$ Université Paris-Cité, CNRS, Astroparticule et Cosmologie, F-75013 Paris, France}
\affiliation{$^{4}$ CNRS-UCB International Research Laboratory, Centre Pierre Bin\'{e}truy, IRL2007, CPB-IN2P3, Berkeley, US}
\affiliation{$^{5}$ INFN Sezione di Ferrara, Via Saragat 1, 44122 Ferrara, Italy}
\affiliation{$^{6}$ Department of Physics, University of Oxford, Denys Wilkinson \\
Building, Keble Road, Oxford OX1 3RH, United Kingdom}
% \setcounter{Maxaffil}{0}
% \renewcommand\Affilfont{\itshape\small}

% \author{Clément Leloup}
%  \altaffiliation[Also at ]{Physics Department, XYZ University.}%Lines break automatically or can be forced with \\
% \author{Josquin Errard}%
% \author{Radek Stompor}
%  \email{Second.Author@institution.edu}
% \affiliation{%
%  Authors' institution and/or address\\
%  This line break forced with \textbackslash\textbackslash
% }%

% \collaboration{MUSO Collaboration}%\noaffiliation

%  \homepage{http://www.Second.institution.edu/~Charlie.Author}
% \affiliation{
%  Second institution and/or address\\
%  This line break forced% with \\
% }%
% \affiliation{
%  Third institution, the second for Charlie Author
% \affiliation{%
%  Authors' institution and/or address\\
%  This line break forced with \textbackslash\textbackslash
% }%

% \collaboration{CLEO Collaboration}%\noaffiliation


\begin{abstract}
    Astrophysical foreground substraction is crucial to retrieve the cosmic microwave background (CMB) polarization out of the observed data. Recent efforts \cite{Leloup:2023vkb,Morshed:2024fow} have been carried out towards the development of a minimally informed component separation method to handle \textit{a priori} unknown
    %tackle inaccurate modeling of the
    foreground spectral energy distributions (SEDs), while being able to estimate both cosmological, foreground, and potentially instrumental parameters, jointly.

    In this paper, we develop a semi-analytical performance forecasting framework for the minimally informed method and
    %and then use it to demonstrate the robustness of the approach.
    %In this paper, we develop a complementary analytical framework to forecast the performance of the method from its bias and statistical error of estimated cosmological parameters.
    we validate it
    %the forecasting tool
    by comparing its results against direct sampling of the harmonic-based likelihood and the pixel domain implementation \texttt{MICMAC}. We then use the forecasting tool to demonstrate the robustness of the bias correction procedure introduced in the minimally informed approach.
    %impact of the accuracy of the fixed CMB covariance estimate, an important ingredient of the minimally informed approach needed to regularize its bias, on the reconstruction of the tensor-to-scalar ratio $r$.
    We find that
    %a highly accurate estimate for the CMB $B$-mode covariance is unnecessary, supporting the applicability of the method in realistic data-analysis context. In particular, an agnostic,
    a data-driven approach based on the currently available observational data is enough to efficiently regularize the bias of the method.
\end{abstract}

\maketitle

\def\matriximg{%
  \begin{matrix}
    \\
    \quad \cdots \quad \\
    \\
   \end{matrix}
}%

\section{Introduction}
\label{section:Introduction}

The observation of the $B$-mode polarization of the Cosmic Microwave Background (CMB) is one of the most important and promising probes for the next stage of cosmological observations. Measuring the amplitude of $B$ modes at large-scale sourced by gravitational waves produced in the early Universe would be a far-reaching discovery that would open an observational window on the popular and successful theoretical paradigm of cosmic inflation. This would constitute the first direct empirical evidence of the fundamentally quantum nature of gravity by accessing energy scales far higher than those accessible from any other probe. As such, the measurement of CMB polarization is the primary scientific goal of current (e.g. BICEP~\cite{BICEP:2021xfz}, the South Pole Telescope \cite{SPT:2023jql,SPT:2025vtb}, Simons Array~\cite{POLARBEAR:2015ixw}, Simons Observatory~\cite{Ade:2018sbj}) and planned future CMB experiments (e.g. LiteBIRD~\cite{Litebird-2019}, CMB-S4~\cite{Abazajian:2016yjj}).

The challenges faced by these experiments in order to retrieve a robust and precise measurement of the large-scale CMB $B$ modes are on par with their cosmological significance. They have to address instrumental difficulties since the cosmological signal is expected to be very faint \cite{Martin:2013tda}, as well as intricate data analysis predicaments, many of which originate from the presence of Galactic foreground emissions at CMB frequencies that are orders of magnitude brighter than the predicted primordial $B$ modes~\cite{Krachmalnicoff:2015xkg}.

Polarized Galactic foregrounds are dominated by synchrotron emissions of charged particles and thermal emissions of dust grain in the interstellar medium, whose exact characteristics are yet to be precisely determined, the best current constraints at CMB frequencies being from Planck observations~\cite{Planck:2018yye}. Because the evolution of foreground emissions with frequency, their Spectral Energy Distributions (SEDs), is known with limited accuracy, multiple models of the Galactic foregrounds spectra with various levels of complexity are still compatible with the available data~\cite{PanEx}. Furthermore, there is now solid evidence that the foreground SEDs are not homogeneous and vary across the sky~\cite{Planck:2015mvg,Krachmalnicoff:2018imw,delaHoz:2023yvz}. %Until accurate physically motivated foreground emission models are available, it is of primary importance to have access to robust statistical methods of foreground cleaning, a process also known as component separation, that are flexible enough to produce robust cosmological results in the widest range possible of realistic sky models.
Until accurate physically motivated foreground emission models are available, it is of primary importance to have access to robust statistical methods of foreground cleaning, the process of removing the foreground contamination from CMB observations, that are flexible enough to produce reliable cosmological results in the widest range possible of realistic sky models.

A large variety of foreground cleaning approaches can be found in the literature, typically split into \textit{parametric} that assume empirical models for the frequency scaling of CMB and foregrounds \cite{Eriksen:2005dr,Stompor:2008sf,delaHoz:2020ggq}, and \textit{non-parametric} methods based on other assumptions such as the Internal Linear Combination (ILC) \cite{WMAP:2003ivt,Eriksen:2004jg,Vio:2008us,Basak:2011yt}, Independent Component Analysis (ICA) \cite{Cardoso:2008qt} or template fitting approaches \cite{Efstathiou:2009MNRAS.397.1355E,Katayama:2011ApJ...737...78K}. Although only \textit{parametric} approaches can perform proper component separation, i.e. recover not only the CMB but also the foreground components, it is customary to use foreground cleaning and component separation interchangeably to refer to all these approaches, as will be done in the following.

As a contribution to addressing this challenge, a new non-parametric method of foreground cleaning, referred to hereafter as the Minimally Informed (MI) component separation, was developed in~\cite{Leloup:2023vkb} with a simple implementation in spherical harmonics domain, later extended to pixel space with the \texttt{MICMAC} package described in~\cite{Morshed:2024fow}. This approach is minimally informed in the sense that it makes only minimal assumptions on the foregrounds, namely the number of independent components, placing it in the same category as ILC. The main difference between these is that the MI approach relies on the extensively studied statistical principle of maximum likelihood instead of the idea of minimum variance used by ILC. The maximum likelihood approach has two main advantages: it is straightforward to extend the likelihood to include instrument parameters for the mitigation of systematic effects, and it renders the estimation and interpretation of results clear and easy to understand, as will be illustrated in the present article. Indeed,
%However, 
studying the performance of a component separation approach in various contexts to assess its robustness by running over a large number of sky and/or instrument simulations may be
%is
very resource intensive in general, but capitalizing on the maximum likelihood estimation framework on which MI is built,
%. As the MI approach is formulated in the extensively studied statistical principle of maximum likelihood,
its performance can be evaluated using simple well-known analytical approximations in the spirit of the two-step framework in~\cite{Stompor:2016hhw} based on the parametric component separation method of~\cite{Stompor:2008sf}. This would grant a better understanding of the behavior of the method itself as well as a mean of study, a better control over its assumptions and a way to explore its interaction with instrumental systematic effects \cite{Ghigna:2020wat,Verges:2020xug,Jost:2022oab,Monelli:2023wmv,LiteBIRD:2023vtc}.

This paper describes such a semi-analytical performance forecasting framework for the study of the MI method in the context of multi-frequency CMB polarization experiments. The general formalism of the forecasting approach is described in Section~\ref{section:General formalism} and an example of case study is presented in Section~\ref{section:Dependence on the correction term} where the performance of the MI method is explored as the characteristics of its regularization scheme change for the search of CMB large-scale $B$ modes with typical ground and space experiments. Finally, the general results of this forecast study are validated using the \texttt{MICMAC} pixel implementation on a small selection of sky simulations, and further discussed in Section~\ref{section:Validation and discussions}.

% In this paper we focus on component separation techniques based on the maximum likelihood principle and discuss the role and impact of the assumptions they rely on. We do so in the context of a unified framework that we develop in Section~\ref{section:General formalism}. Then, we capitalize on the obtained insights and propose a new maximum likelihood-based, non-parametric approach to foreground cleaning with a minimal set of assumptions on foreground components. In Section~\ref{section:Implementation of the spectral likelihood}, we detail a simple implementation of the proposed method in harmonic domain, that we use on study cases relevant for future experiments in Section~\ref{section:Performances of the method}.

\section{General formalism}
\label{section:General formalism}

\subsection{Minimally Informed Component Separation}

\subsubsection{Data model}

We suppose the observed data can be summarized in a (frequency $\nu$) $\times$ (pixel $p$) data vector $\mathbf{d}$ that is a frequency dependent linear combination of $n_{c}$ component templates $\mathbf{s}$ and random noise $\mathbf{n}$,
\begin{equation}
    \mathbf{d} = \mathbf{Bs} + \mathbf{n},
\end{equation}
where we introduced the mixing matrix $\mathbf{B}$ whose elements $B_{c}^{\nu}$ is the coefficients of the component $c$ in the linear combination at frequency $\nu$. Following \cite{Leloup:2023vkb}, we identify the CMB component $\mathbf{s_c} = \mathbf{Es}$, where $\mathbf{E}$ is an operator that selects only the CMB component, by imposing the elements of the CMB column of $\mathbf{B}$ to be fixed to 1 (in CMB units) as well as a Gaussian prior on $s_{c}$. In the absence of further assumptions, the definition of the foreground components is ambiguous as the observed data is invariant under the transformation
\begin{eqnarray}
    && \mathbf{B} \rightarrow \mathbf{B'} = \mathbf{B} \left[ \begin{array}{c|c}
  		\rule[-1.2ex]{0pt}{0pt} 1 & \mathbf{0}_{n_{c}-1}^{T} \\
		\hline
		\rule{0pt}{2.5ex} \mathbf{0}_{n_{c}-1} & \mathbf{B}_\mathbf{f}
	\end{array} \right]^{-1} \equiv \mathbf{BM}^{-1}, \\
    && \mathbf{s} \rightarrow \mathbf{s'} = \mathbf{Ms}.
\end{eqnarray}

In the above expression, we defined $\mathbf{0}_{n}$ to be a vector of size $n$ filled with zeros. To lift this degeneracy, we fix some of the remaining free elements of $\mathbf{B}$ and take it to be of the form
\begin{eqnarray}
    \mathbf{B} = \left[ \begin{array}{c|c}
		 & \rule[-1.2ex]{0pt}{0pt} { \mathbf{b_f}}^{T} \\ \cline{2-2}
		 & \\
		 & \mathbb{1}_{n_{c}-1} \\
		 \mathbf{B_{c}} & \\ \cline{2-2}
		 & \\
		 & \mathbf{B_{f}} \\
		 & \\
	\end{array}	 \right], \\
    \nonumber
\end{eqnarray}
where we have defined the CMB part of the mixing matrix $\mathbf{B_{c}}$, the identity matrix of size $n \times n$, $\mathbb{1}_{n}$, and where $\mathbf{b_{f}}$ and $\mathbf{B_{f}}$ are a vector and a rectangular matrix of free parameters respectively. This means that each foreground component in the MI approach is not identified from their physical characteristics, but is instead defined as the non-CMB contribution in a selected frequency channel.

\subsubsection{The likelihood}

The extraction of CMB from the foreground contaminated frequency maps is performed in this method by using a maximum likelihood approach. The likelihood is obtained by first assuming that the noise $\mathbf{n}$ is a multivariate Gaussian with covariance matrix $\mathbf{N}$ and, as explained in the previous section, that we have a Gaussian prior on the CMB signal component with covariance matrix $\mathbf{C}$,
\begin{eqnarray}
    \mathcal{S} \left( \mathbf{s}, \beta, \gamma \right) & \equiv & -2\, \mathrm{ln}\, \mathcal{L} \nonumber \\
    & = & \mathrm{const} + \left( \mathbf{d} - \mathbf{B} \left( \beta \right) \mathbf{s} \right)^{T} \mathbf{N}^{-1} \left( \mathbf{d} - \mathbf{B} \left( \beta \right) \mathbf{s} \right) \nonumber \\
    && +\, \mathbf{s_{c}}^{T} \left( \mathbf{C} \left( \gamma \right) \right)^{-1} \mathbf{s_{c}} + \mathrm{ln} \left| \mathbf{C} \left( \gamma \right) \right|,
\end{eqnarray}
where $\left| \, \cdot \, \right|$ stands for the determinant, and $\beta$ and $\gamma$ are the free parameters of the mixing matrix and CMB covariance respectively. As the constant terms have no impact on the results, we stop including them in the following. We can get rid of the dependence on the signal components $\mathbf{s}$ and $\mathbf{s_{c}}$ by maximizing first over the foreground components, then marginalizing over the CMB component to obtain the spectral likelihood
\begin{eqnarray}
    \mathcal{S}_{\rm spec} \left( \beta, \gamma \right) & = & \mathbf{d}^{T}\mathbf{Pd} + \hatsc^{T} \left( \mathbf{C} + \Nc \right)^{-1} \hatsc \nonumber \\
    && +\, \mathrm{ln} \left| \mathbf{C} + \Nc \right| - \mathrm{ln} \left| \Nc \right|. \label{eq:uncorrected likelihood}
\end{eqnarray}

In the expression of the spectral likelihood above, we introduced the projection operator $\mathbf{P}$ on the space orthogonal to the columns of $\mathbf{B}$, the estimated CMB signal component $\hatsc$ and the noise after component separation $\Nc$, defined respectively by
\begin{eqnarray}
    \mathbf{P} & \equiv & \mathbf{N}^{-1} - \mathbf{N}^{-1}\mathbf{B} \left( \mathbf{B}^{T}\mathbf{N}^{-1}\mathbf{B} \right)^{-1} \mathbf{B}^{T}\mathbf{N}^{-1}, \label{eq:projection operator} \\
    \Nc & \equiv & \mathbf{E}^{T} \left( \mathbf{B}^{T} \mathbf{N}^{-1} \mathbf{B} \right)^{-1} \mathbf{E}, \label{eq:Nc} \\
    \hatsc & \equiv & \mathbf{E}^{T} \left( \mathbf{B}^{T} \mathbf{N}^{-1} \mathbf{B} \right)^{-1} \mathbf{B}^{T} \mathbf{N}^{-1} \mathbf{d} \equiv \mathbf{Wd}. \label{eq:hat sc}
\end{eqnarray}

In the last equality above, we introduced the weighting operator $\mathbf{W}$ that linearly combines the frequency dependent observed data to produce the CMB estimated signal $\hatsc$.

It has, however, been shown in \cite{Leloup:2023vkb} that the likelihood \eqref{eq:uncorrected likelihood} is plagued by a significant bias and need to be regularized. In the present work, we consider a different bias correction scheme than the one used in \cite{Leloup:2023vkb}, that was initially proposed in \cite{Morshed:2024fow} and which consists in introducing an ad-hoc correction to the likelihood that now reads
\begin{eqnarray}
    \mathcal{S}_{\rm MI} & = & \mathbf{d}^{T}\mathbf{Pd} + \hatsc^{T} \left( \mathbf{C} + \Nc \right)^{-1} \hatsc \nonumber \\
    && +\, \mathrm{ln} \left| \mathbf{C} + \Nc \right| - \mathrm{ln} \left| \Ctilde + \Nc \right|, \label{eq:MI likelihood}
\end{eqnarray}
where $\Ctilde$ is a fixed estimate of the CMB covariance matrix. The main goal of the present work is to study the impact of this estimate on the ability to reconstruct the foreground and CMB parameters.

\subsection{Averaged formalism}

We are mostly interested in two quantities: the bias on the cosmological parameters that enter the parameterization of $\mathbf{C}$ and their standard deviation that give the related statistical errors. %As we are interested in general statements, but generating multiple noise and CMB simulations as well as running the component separation for all relevant cases of the CMB estimate $\Ctilde$ in the correction term
As generating and running the component separation over multiple a large number of noise and CMB simulations
would be too resource demanding, we adopt a simplified approximate approach following a Fisher-like formalism based on the ensemble averaged likelihood that enables us to reconstruct these two quantities. The component separation formulation as a problem of likelihood maximization allows us to take the noise and CMB average analytically from the likelihood expression. This is easily done by noting that the likelihood \eqref{eq:MI likelihood} can be rephrased in such a way that the data enters the only as a quadratic contribution $\mathbf{dd}^{T}$, which means that the noise and CMB average of the likelihood only depends on its variance:
\begin{equation}
    \left< \mathbf{dd}^{T} \right> = \left< \left( \mathbf{Bs} + \mathbf{n} \right) \left( \mathbf{Bs} + \mathbf{n} \right)^{T} \right> \equiv \mathbf{D} + \mathbf{N}, \label{eq:data covariance}
\end{equation}
where brackets indicate taking the ensemble average over noise and CMB realizations. To get the second equality, we assumed no correlation between the noise $\mathbf{n}$ and the astrophysical signal $\mathbf{s}$ which is a weak assumption, and we introduced the variance of the noiseless data $\mathbf{D}$. In simple cases when the data model actually matches the observed sky, it is expressed from the foreground map templates, the true mixing matrix $\mathbf{\hat{B}}$ and the true CMB covariance $\mathbf{\hat{C}}$ by:
\begin{equation}
    \mathbf{D} = \mathbf{\hat{B}} \left[ \begin{array}{c|c}
        \rule[-1.2ex]{0pt}{0pt} \mathbf{\hat{C}} & \mathbf{0}_{n_{c}-1}^{T} \\ \hline
        \rule{0pt}{2.5ex} \mathbf{0}_{n_{c}-1} & \mathbf{s_{f}s_{f}}^{T}
    \end{array} \right] \mathbf{\hat{B}}^{T}.
\end{equation}

However, keeping it general as $\mathbf{D}$ allows for the study of cases where there is a mismatch between these as can happen in the presence of systematic effects unaccounted for, for instance.

From this data covariance, we find a simple expression for the noise and CMB averaged likelihood of the MI method,
\begin{eqnarray}
    \left< \mathcal{S}_{\rm MI} \right> & = & \mathrm{Tr} \left[ \mathbf{P} \mathbf{D} - \left( \mathbf{C} + \Nc \right)^{-1} \hatD \right] \nonumber \\
    && +\, \mathrm{ln} \left| \mathbf{C} + \Nc \right| - \mathrm{ln} \left| \mathbf{\tilde{C}} + \Nc \right|, \label{eq:MI likelihood avg}
\end{eqnarray}
with the quantity $\hatD$ being the difference between the parameterized CMB covariance and the true covariance in the reconstructed CMB map after component separation,
\begin{equation}
    \hatD \equiv \mathbf{C} - \mathbf{W} \mathbf{D} \mathbf{W}^{T}.
\end{equation}

If the data model matches the actual observed sky and if all the parameters of the likelihood are at their true value, i.e. $\mathbf{B} = \mathbf{\hat{B}}$ and $\mathbf{C} = \mathbf{\hat{C}}$, $\hatD$ vanishes. Because the cosmological parameters directly enter the parameterization of $\mathbf{C}$ and, hence, of the likelihood, the bias is simply found as the difference between their value at the peak of the likelihood and their true values,
\begin{equation}
    \delta \gamma = \left( \argmax_{\theta} \left< \mathcal{L}_{\rm MI} \right> \right)_{\gamma} - \gamma_{\rm true}, \label{eq:bias definition}
\end{equation}
where $\theta$ is the concatenated vector of all the free parameters $\theta = \left( \beta, \gamma \right)$. In our approach, the value at the peak is obtained by finding the parameters that make the gradient of the likelihood vanish. This is enough in the case we considered as the likelihood \eqref{eq:MI likelihood avg} is sufficiently well-behaved, i.e. there is no saddle point. Therefore, this requires to compute the first derivatives of the averaged likelihood with respect to all the free parameters that enter the likelihood. They are given by,
\begin{widetext}
\begin{eqnarray}
    && \left< \frac{\partial \mathcal{S}_{\rm MI}}{\partial \beta} \right> = \mathrm{Tr} \left[ \left( \mathbf{C} + \Nc \right)^{-1} \Nc_{, \beta} \left( \mathbf{C} + \Nc \right)^{-1} \hatD - 2\, \mathbf{N}^{-1} \mathbf{B} \left( \mathbf{B}^{T} \mathbf{N}^{-1} \mathbf{B} \right)^{-1} \mathbf{B}_{, \beta}^{T} \mathbf{P} \mathbf{D} \right. \nonumber \\
    && \left. - \left( \mathbf{C} + \Nc \right)^{-1} \hatD_{, \beta} + \left( \mathbf{C} + \Nc \right)^{-1} \Nc_{, \beta} - \left( \mathbf{\tilde{C}} + \Nc \right)^{-1} \Nc_{, \beta} \right], \label{eq:dS/dbeta} \\ [0.2cm]
    && \left< \frac{\partial \mathcal{S}_{\rm MI}}{\partial \gamma} \right> = \mathrm{Tr} \left[ \left( \mathbf{C} + \Nc \right)^{-1} \mathbf{C}_{, \gamma} \left( \mathbf{C} + \Nc \right)^{-1} \hatD \right]. \label{eq:dS/dgamma}
\end{eqnarray}
\end{widetext}

In the above expressions and in the following, a comma subscript means we are taking the derivative of the operator with respect to the subsequent parameter.

In order to get an estimate of the statistical errors of the parameter once the maximum of the likelihood is found, we assume that the averaged likelihood is Gaussian around its peak. In this case, the covariance matrix between the free parameters is given by the Hessian of the negative averaged log-likelihood, $- \left< \mathrm{ln}\, \mathcal{L} \right>$, evaluated at its peak. Although not evaluated at the true value of the parameters, it will be referred to in the following as the Fisher information matrix $\mathcal{F}$. The estimate of the statistical error is defined from the inverse of the Fisher matrix following
\begin{equation}
    \sigma_{\gamma} = \sqrt{\left( \mathcal{F}^{-1} \right)_{\gamma \gamma}}. \label{eq:sigma from Fisher}
\end{equation}

We know that the full likelihood can not be Gaussian as the derivative of the averaged likelihood with respect to the parameters of $\mathbf{C}$ (Eq.~\eqref{eq:dS/dgamma}) vanishes when the parameters are at their true value while the derivatives with respect to the mixing matrix elements (Eq.~\eqref{eq:dS/dbeta}) is non-zero, which would be impossible for a Gaussian likelihood. However, as we will see in Section~\ref{section:Dependence on the correction term}, the likelihood is sufficiently close to Gaussian in the cases under study. Thus, we need the analytical expressions of the second derivatives of the noise and CMB averaged likelihood which are,
\begin{widetext}
\begin{eqnarray}
    && \left< \frac{\partial^{2} \mathcal{S}_{\rm MI}}{\partial \beta' \partial \beta} \right> = \mathrm{Tr} \left[ 2 \left\{ \mathbf{N}^{-1} \mathbf{B} \left( \mathbf{B}^{T} \mathbf{N}^{-1} \mathbf{B} \right)^{-1} \mathbf{B}_{, \beta}^{T} \mathbf{N}^{-1} \mathbf{B} \left( \mathbf{B}^{T} \mathbf{N}^{-1} \mathbf{B} \right)^{-1} \mathbf{B}_{, \beta'}^{T} \mathbf{P} \right. \right. \nonumber \\
    && + \mathbf{N}^{-1} \mathbf{B} \left( \mathbf{B}^{T} \mathbf{N}^{-1} \mathbf{B} \right)^{-1} \mathbf{B}_{, \beta'}^{T} \mathbf{N}^{-1} \mathbf{B} \left( \mathbf{B}^{T} \mathbf{N}^{-1} \mathbf{B} \right)^{-1} \mathbf{B}_{, \beta}^{T} \mathbf{P} - \mathbf{P} \mathbf{B}_{, \beta'} \left( \mathbf{B}^{T} \mathbf{N}^{-1} \mathbf{B} \right)^{-1} \mathbf{B}_{, \beta}^{T} \mathbf{P} \nonumber \\
    && \left. \left. + \mathbf{N}^{-1} \mathbf{B} \left( \mathbf{B}^{T} \mathbf{N}^{-1} \mathbf{B} \right)^{-1} \mathbf{B}_{, \beta}^{T} \mathbf{P} \mathbf{B}_{, \beta'} \left( \mathbf{B}^{T} \mathbf{N}^{-1} \mathbf{B} \right)^{-1} \mathbf{B}^{T} \mathbf{P} - \mathbf{N}^{-1} \mathbf{B} \left( \mathbf{B}^{T} \mathbf{N}^{-1} \mathbf{B} \right)^{-1} \mathbf{B}_{, \beta \beta'}^{T} \mathbf{P} \right\} \mathbf{D} \right. \nonumber \\
    && - 2 \left( \mathbf{C} + \Nc \right)^{-1} \Nc_{, \beta'} \left( \mathbf{C} + \Nc \right)^{-1} \Nc_{, \beta} \left( \mathbf{C} + \Nc \right)^{-1} \hatD + \left( \mathbf{C} + \Nc \right)^{-1} \Nc_{, \beta\beta'} \left( \mathbf{C} + \Nc \right)^{-1} \hatD \nonumber \\
    && + \left( \mathbf{C} + \Nc \right)^{-1} \Nc_{, \beta} \left( \mathbf{C} + \Nc \right)^{-1} \hatD_{, \beta'} + \left( \mathbf{C} + \Nc \right)^{-1} \Nc_{, \beta'} \left( \mathbf{C} + \Nc \right)^{-1} \hatD_{, \beta} - \left( \mathbf{C} + \Nc \right)^{-1} \hatD_{, \beta \beta'} \nonumber \\
    && + \left( \mathbf{C} + \Nc \right)^{-1} \Nc_{, \beta\beta'} - \left( \mathbf{C} + \Nc \right)^{-1} \Nc_{, \beta'} \left( \mathbf{C} + \Nc \right)^{-1} \Nc_{, \beta} \nonumber \\
    && \left. - \left( \mathbf{\tilde{C}} + \Nc \right)^{-1} \Nc_{, \beta\beta'} + \left( \mathbf{\tilde{C}} + \Nc \right)^{-1} \Nc_{, \beta'} \left( \mathbf{\tilde{C}} + \Nc \right)^{-1} \Nc_{, \beta} \right], \label{eq:d2S/db2} \\ [0.2cm]
    && \left< \frac{\partial^{2} \mathcal{S}_{\rm MI}}{\partial \beta \partial \gamma} \right> = 2 \mathrm{Tr} \left[ \left( \mathbf{C} + \Nc \right)^{-1} \mathbf{C}_{, \gamma} \left( \mathbf{C} + \Nc \right)^{-1} \left( \hatD_{, \beta} - \hatD \left( \mathbf{C} + \Nc \right)^{-1} \Nc_{, \beta} \right) \right], \label{eq:d2S/dbdgamma} \\ [0.2cm]
    && \left< \frac{\partial^{2} \mathcal{S}_{\rm MI}}{\partial \gamma^{2}} \right> = \mathrm{Tr} \left[ \left( \mathbf{C} + \Nc \right)^{-1} \left( \mathbf{C}_{, \gamma \gamma} - 2\mathbf{C}_{, \gamma} \left( \mathbf{C} + \Nc \right)^{-1} \mathbf{C}_{, \gamma} \right) \left( \mathbf{C} + \Nc \right)^{-1} \hatD \right. \nonumber \\
    && \left. + \left( \mathbf{C} + \Nc \right)^{-1} \mathbf{C}_{, \gamma} \left( \mathbf{C} + \Nc \right)^{-1} \mathbf{C}_{, \gamma} \right]. \label{eq:d2S/dgamma2}
\end{eqnarray}
\end{widetext}

From our parameterization of the mixing matrix using its elements directly, the second derivatives of $\mathbf{B}$ all vanish. The expression of the first and second derivatives of $\Nc$ and $\hatD$ with respect to $\beta$ are given in Appendix~\ref{appendix:Derivatives entering the Fisher information matrix}. We summarize the procedure that allows us to estimate the bias $\delta \gamma$ and statistical errors $\sigma_{\gamma}$ for the cosmological parameters $\gamma$:
\begin{enumerate}
    \item We define the noiseless data covariance operator $\mathbf{D}$ from simulations of the observed sky as implicitly defined in Eq.~\eqref{eq:data covariance}.
    \item We explore the parameter space of the likelihood to find its maximum as the position where all of the derivatives of Eq.~\eqref{eq:dS/dbeta}-\eqref{eq:dS/dgamma} vanish and recover the bias $\delta \gamma$ as the distance between their value at the peak of the likelihood and their true values, as in Eq.~\eqref{eq:bias definition}.
    \item We compute the Fisher information matrix $\mathbf{F}$ that corresponds to the Hessian of the negative averaged log-likelihood at the likelihood maximum found in step 2. from Eq.~\eqref{eq:d2S/db2}-\eqref{eq:d2S/dbdgamma}. The estimate of the statistical error $\sigma_{\gamma}$ is found from its inverse using Eq.~\eqref{eq:sigma from Fisher}.
    \item Results from the two previous steps are then used to run extensive tests in order to extract general results on the performance of the method. These are eventually validated by running the \texttt{MICMAC} package on a small set of selected simulations.
\end{enumerate}

\begin{figure*}[!htb]
\begin{center}
\includegraphics[width=1.5\columnwidth]{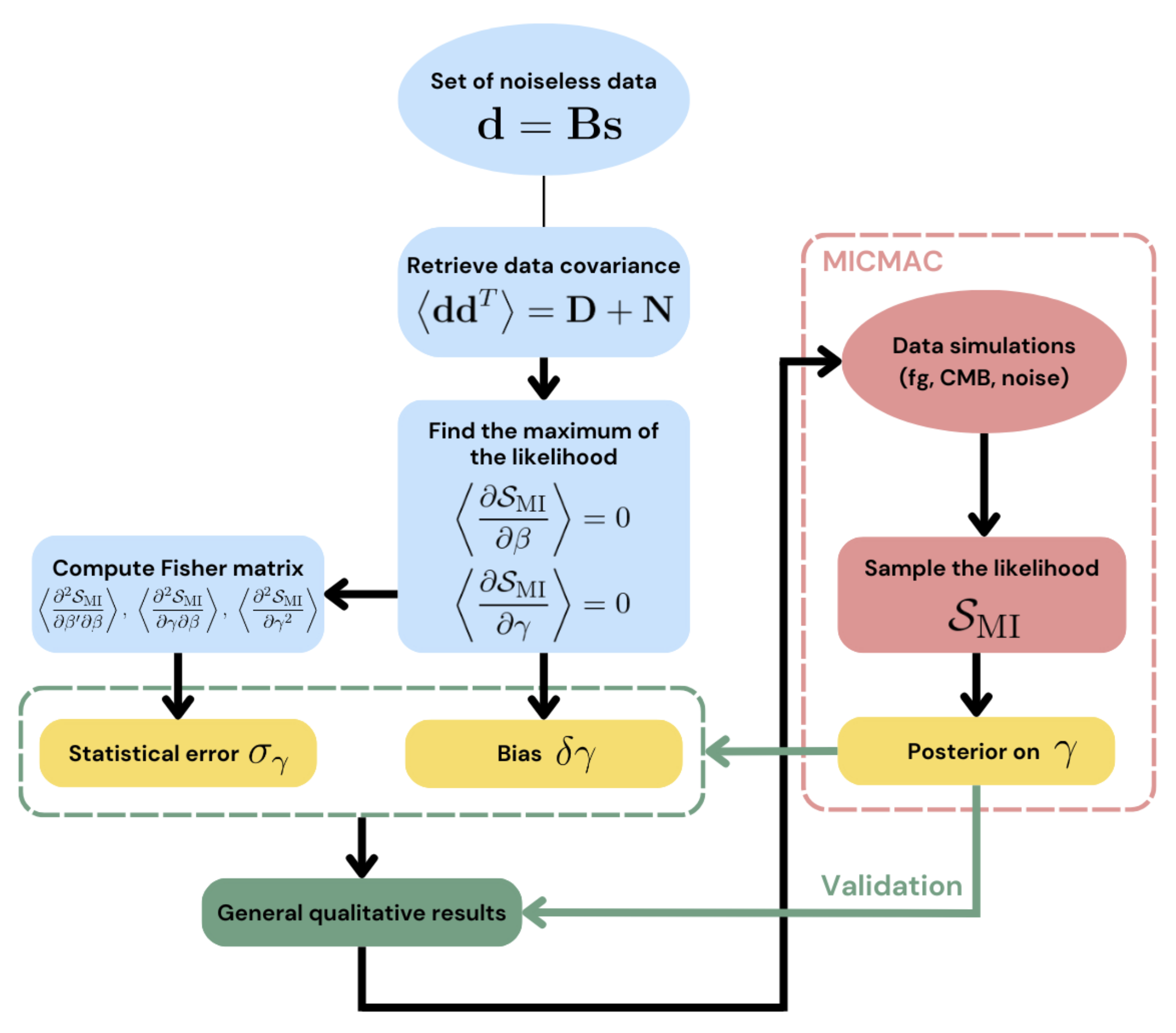}
\caption{Schematic view of the different steps of a typical analysis involving the use of the simplified Fisher forecast approach developed in Section~\ref{section:General formalism}, including the validation of the results using the \texttt{MICMAC} package which directly samples the likelihood in Eq.~(\ref{eq:MI likelihood}). It is, in particular, used in the study of the correction term presented in Section~\ref{section:Dependence on the correction term}.}
%https://www.canva.com/design/DAGShBWHkdE/Yu4joJVEiLO_w_vMdKSwXw/edit?utm_content=DAGShBWHkdE&utm_campaign=designshare&utm_medium=link2&utm_source=sharebutton
\label{fig:analysis pipeline}
\end{center}
\end{figure*}

The analysis pipeline described above and used in Section~\ref{section:Dependence on the correction term} is illustrated in FIG.~\ref{fig:analysis pipeline}.

Although this procedure is very simple and builds on analytical expressions, the matrix algebra is only tractable when working in harmonic space which requires foregrounds with spatially homogeneous scaling as well as homogeneous noise. This is a strong limitation of the framework which is therefore not able to give quantitative results regarding the performance of the method to complex foregrounds. However, it is well suited to study its robustness to effects that can be decoupled from the complexity of the foregrounds such as systematic effects as is still done with various other component separation methods, see e.g. \cite{Ghigna:2020wat,Verges:2020xug,Jost:2022oab,Monelli:2023wmv,LiteBIRD:2023vtc}. In addition, it is also not impossible that the approach followed in \cite{Errard:2018ctl} to analytically estimate the foreground leakage when using a spatially variable component separation from the homogeneous case could be adapted to the present approach and give a quantitative estimate of $\delta r$ and $\sigma_{r}$ even when dealing with spatially varying foregrounds. In the meantime, the general results derived from the framework presented here will be further validated by running the pixel version implemented in the \texttt{MICMAC} package on specific noise and CMB realizations.

As an application of the procedure, and an interesting study in itself, we will use this formalism in the next section to explore the robustness of the minimally informed component separation to change in its correction term $\mathbf{\tilde{C}}$.

\section{Dependence on the correction term}
\label{section:Dependence on the correction term}

\subsection{Implementation of the formalism}
\label{subsection:Implementation of the formalism}

We present here the specific implementation of the formalism presented in Section~\ref{section:General formalism}. As previously mentioned, the approach needs to be used in the harmonic domain in order to be tractable in a short time. Working in the harmonic domain is also convenient as the covariance of the CMB signal is best expressed as its angular power-spectrum. As most future CMB experiments will focus on the measurement of its polarization, i.e. $Q$ and $U$ Stokes parameters of the pixelized micro-wave sky which correspond to $E$ and $B$ modes in the harmonic domain, we will focus on these and will not consider the CMB intensity $I$. Although evidence of different SEDs for the $Q$ and $U$ Stokes parameters has been found in data \cite{Pelgrims:2021gqi}
%it is expected that the $Q$ and $U$ SEDs differ slightly
due to averaging of varying foregrounds along the line of sight \cite{Tassis:2014bea,Poh:2016rfa}, for simplicity and because it is still the standard approach for the analysis of available data \cite{Planck:2018yye,delaHoz:2023yvz} given the small size of the effect, we assume these SEDs to be the same which allows us to
%For simplicity and because there is no conclusive evidence that the $Q$ and $U$ SEDs are different, we
use the same mixing matrix for both $E$ and $B$ modes. The CMB $E$ and $B$-mode signal covariance in harmonic space is their angular power-spectrum, which we evaluate at multipoles up to $\ell < \ell_{\rm max} = 256$ and with minimum multipole constrained by the sky coverage of the experimental set-ups.

Because the main scientific goal of upcoming CMB experiments is the measurement of large-scale $B$ modes, where the cosmological signal relevant for cosmic inflation is to be found, only the CMB primordial $B$ mode power-spectrum will be fitted for, the contributions in the $E$ modes and the lensing part of the $B$ modes will be assumed to be purely Gaussian and perfectly known. The impact of an imperfect knowledge of these, of non-Gaussianities of the lensing $B$-modes as well as the capabilities of the methodology to recover them is left for future work. The theoretical spectra are obtained with \texttt{CAMB}~\cite{Lewis:1999bs} using the cosmological parameters from the Planck 2018 cosmological analysis best-fit \cite{Planck:2018vyg}.
%and are particularly interested in the measurements of large-scale $B$ modes, where the cosmological signal relevant for cosmic inflation is to be found. Therefore, we focus only on the CMB $B$-mode angular power spectrum at multipoles $\ell < \ell_{\rm max} = 256$, discarding information from intensity and $E$ modes of polarization.
The multipole-dependent CMB $B$-mode signal covariance used in the likelihood \eqref{eq:MI likelihood avg} is parameterized as
\begin{equation}
    \left( \mathbf{C}^{BB} \right)_{\ell} \left( r \right) = C_{\ell}^{BB} \left( r \right) =  r C_{\ell}^{BB, \rm prim} + C_{\ell}^{BB, \rm lens}. \label{eq:CMB power-spectrum}
\end{equation}

\begin{table}[!htb]
\renewcommand{\arraystretch}{2}
\begin{center}
\begin{tabular}{|P{2cm}|c|c|c|c|c|c|}
\hline
\multicolumn{7}{|c|}{Ground experiment} \\ \hline\hline
\multirow{1}{2cm}{\centering $\nu$\\ (GHz)} & 27 & 39 & 93 & 145 & 225 & 280 \\ \hline
\multirow{1}{2cm}{\centering Sensitivity\\ ($\mu$K-arcmin)} & 49.5 & 29.7 & 3.7 & 4.7 & 8.9 & 22.6 \\ \hline
$f_{\rm sky}$ & \multicolumn{6}{c|}{$\sim 10\%$} \\ \hline
\end{tabular}
\caption{Principal characteristics of the ground experiment set-up used in this study: center frequency of the frequency bands, sensitivity in polarization and observed fraction of the sky corresponding to the approximate sky coverage of the SO-SAT \cite{SimonsObservatory:2018koc}.}
\label{tab:ground experiment}
\end{center}
\end{table}

Here $C_{\ell}^{BB, \rm prim}$ is the primordial contribution to $B$ modes, $C_{\ell}^{BB, \rm lens}$ is the contribution from gravitationally lensed E modes computed with the \texttt{CAMB} code \cite{Camb} using Planck parameters \cite{Planck:2018vyg}, and $r$ is the tensor-to-scalar ratio which is assumed to be the only free parameter of the CMB covariance for this study, whose bias and statistical error constitute our main figures of merit. Following the assumptions in \cite{Leloup:2023vkb} and \cite{Morshed:2024fow}, we use the lensing power-spectrum $C_{\ell}^{BB, \rm lens}$ as our baseline for the estimated CMB $B$-mode covariance $\left( \Ctilde^{BB} \right)_{\ell}$ and study how the performance of the method changes around this baseline. As it would not make sense to explore the impact of varying the estimated CMB $E$-mode covariance $\Ctilde^{EE}$ without adjusting for it, we assume $\left( \Ctilde^{EE} \right)_{\ell}$ to be the theoretical $E$-mode power-spectrum used to produce the CMB simulated data.

Similarly to the CMB covariance, the noise covariance is given by its angular power spectrum $\mathbf{N}_{\ell}$, and we assume a frequency-channel dependent white noise in the following. The level of this white noise is determined by the experimental set-up under consideration. The specific set-ups we consider correspond to a ground based experiment, with characteristics based on the SO-SATs \cite{SimonsObservatory:2018koc} that we detail in TABLE~\ref{tab:ground experiment}, and a space-borne experiment with properties based on LiteBIRD \cite{LiteBIRD:2022cnt} detailed in TABLE~\ref{tab:space experiment}.

\begin{table}[!htb]
\renewcommand{\arraystretch}{2}
\begin{adjustbox}{width=\columnwidth,center}
\begin{tabular}{|c|c||c|c|}
\hline
\multicolumn{4}{|c|}{Space-borne experiment} \\ \hline\hline
\multirow{1}{2cm}{\centering $\nu$\\ (GHz)} & \multirow{1}{2cm}{\centering Sensitivity\\ ($\mu$K-arcmin)} & \multirow{1}{2cm}{\centering $\nu$\\ (GHz)} & \multirow{1}{2cm}{\centering Sensitivity\\ ($\mu$K-arcmin)} \\ \hline
40 & 37.42 & 140 & 4.79 \\ \hline
50 & 33.46 & 166 & 5.57 \\ \hline
60 & 21.31 & 195 & 5.85 \\ \hline
68 & 16.87 & 235 & 10.79 \\ \hline
78 & 12.07 & 280 & 13.80 \\ \hline
89 & 11.3 & 337 & 21.95 \\ \hline
100 & 6.56 & 402 & 47.45 \\ \hline
119 & 4.58 \\ \hline
$f_{\rm sky}$ & \multicolumn{3}{c|}{$\sim 50\%$} \\ \hline
\end{tabular}
\end{adjustbox}
\caption{Characteristics of the space-borne experiment set-up used in this study: center frequency of the frequency bands, sensitivity in polarization and observed fraction of the sky. Although a space-borne experiment has access to the entirety of the sky, we use the apodized Planck HFI 60\% mask \cite{PLA} to hide the Galactic plane, leaving an effective sky fraction of approximately half of the sky.}
\label{tab:space experiment}
\end{table}

It is well-known that spherical harmonic transform defined on an incomplete sky signal need to be treated with utmost care as it requires substantial effort to mitigate mixing between $E$ and $B$ modes that could potentially interfere with the component separation procedure. As a proper treatment can be challenging to implement in practice and because it is not the focus of this work, we assume here that the only effect of the incomplete sky coverage is to reduce the number of modes accessible to observation, thus increasing the statistical uncertainties on the parameters and rendering the lowest multipoles entirely unconstrained. This overall reduction of power is taken into account by including an overall global factor of $f_{\rm sky}$ in the likelihood of Eq.~\eqref{eq:MI likelihood avg}, that appears from the unobserved $m$ multipoles at a given $\ell$. While this approximation has an impact on the statistical uncertainties, we do not expect the bias recovered using the average likelihood to be affected because it appears as a local value of $\gamma$ in the observed patch of the sky, but this induced bias should vanish when averaging over CMB and noise realizations. %Thus, the results using the averaged likelihood should not be affected.
In addition, the fraction of the sky observable by the experiment determines, in addition to the filtering of data observed from the ground that we do not cover here, the lowest multipole accessible from the observed data. In the following we will be using multipoles in the range $30 \leq \ell \leq 256$ for the Ground experiment, and $2 \leq \ell \leq 256$ for the Space experiment. Besides, since the foreground model considered does not feature spatial variability of its SEDs, the exact location of the observed area will not have a significant impact on the ability to clean it. In summary, in the present study, the impact of the incomplete sky coverage is entirely described by two numbers, $f_{\rm sky}$ and the lowest multipole, without need to specify in details the observed footprint. The correct way to account for the presence of the mask is to follow the implementation in \texttt{MICMAC} \cite{Morshed:2024fow}, and it has been verified that the approximation described above gives consistent results for the cases under study.

%The fraction of the sky observable by the experiment determines the lowest multipole accessible from the observed data. In the following we will be using multipoles in the range $30 \leq \ell \leq 256$ for the Ground experiment, and $2 \leq \ell \leq 256$ for the Space experiment. In addition, the fraction of the sky observed will affect the statistical uncertainties on the parameters as less degrees of freedom are accessible to observation. This reduction of power is taken into account by including an overall global factor of $f_{\rm sky}$ in the likelihood \eqref{eq:MI likelihood avg}, that appears from the unobserved $m$ multipoles at a given $\ell$. This allows us to stay in the domain of spherical harmonics coefficient and to avoid any $E$-to-$B$ leakage that would arise from an overly simplistic masking procedure. This approximation only affects the statistical uncertainties and not the bias that appears as a local value of $\gamma$ in the observed patch of the sky, but this induced bias should vanish when averaging over CMB and noise realizations. Thus, the results using the averaged likelihood should not be affected.
%It has been verified by comparing with the pixel implementation in \texttt{MICMAC} \cite{Morshed:2024fow} which properly takes the effect of the mask into account, that the approximation described here gives consistent results.

From the current knowledge of the polarized foregrounds in the frequency range under consideration, see e.g. \cite{Planck:2018yye}, we assume the presence of two polarized foreground components, therefore assuming a three-column mixing matrix. The foreground templates with their spatially homogeneous SED are taken from the \texttt{d0} model for Galactic thermal dust and the \texttt{s0} model for synchrotron emission of the \texttt{PySM} software \cite{Thorne:2016ifb,Zonca:2021row}. This choice is motivated by our implementation in the space of spherical harmonic coefficients which is not suited to handle more complex foregrounds with spatial variations of their SEDs. The true CMB covariance $\mathbf{\hat{C}}$ used in the definition of the data covariance $\mathbf{D}$ is taken to have the same form as Eq.~\eqref{eq:CMB power-spectrum} with values of the tensor-to-scalar ratio $r$ specified for each case.

\begin{figure*}[!htb]
\begin{center}
\includegraphics[width=2.0\columnwidth]{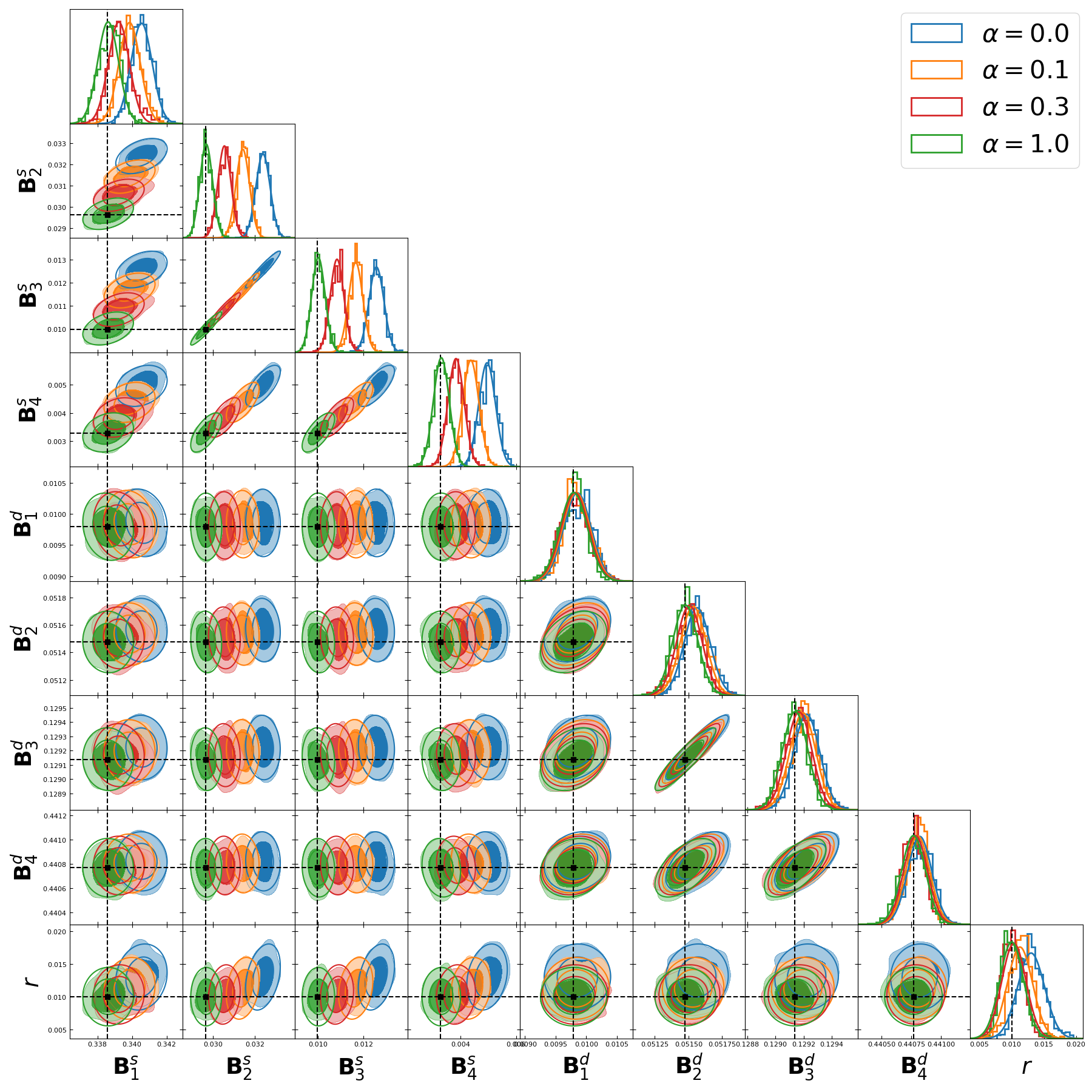}
\caption{2D and 1D densities for the free parameters of the likelihoods in Eq.~\eqref{eq:MI likelihood avg}, i.e. the free elements of the mixing matrix $\mathbf{B}$ and the tensor-to-scalar ratio $r$, for the ground experiment case and for various values of the amplitude $\alpha$ of the CMB covariance estimate (see Eq.~\eqref{eq:Ctilde alpha}). The 2D filled contours are obtained by directly sampling the likelihood \eqref{eq:MI likelihood avg} including $\sim 1000$ independent samples for each case and are plotted using the \texttt{GetDist} package \cite{Lewis:2019xzd}, while the ellipses are calculated from the bias and Fisher information matrix defined in our simplified formalism. Similarly, the 1D histograms are from the sampling while the Gaussian curves are from the simplified formalism. The mixing matrix elements are displayed in the format $\mathbf{B}_{\nu}^{c}$, where $\nu \in 0 \hdots 5$ are the frequency channels and $c \in s,d$ correspond to the foreground component ($s$ for synchrotron and $d$ for dust). The black points and dashed lines correspond to the true value of the parameters used to produce the input signal. In all cases, the model of foregrounds used is \texttt{d0s0}.}
\label{fig:comparison sampling vs forecast}
\end{center}
\end{figure*}

\subsection{Validation of the forecasting tool}

We start our study by validating the accuracy of the forecasting approach described in Section~\ref{section:General formalism}. This validation is performed by comparing its results with those obtained from the direct sampling of the averaged likelihood \eqref{eq:MI likelihood avg} with the Monte-Carlo Markov Chain (MCMC) package \texttt{emcee} \cite{emcee}, for a range of cases from small to large bias on $r$. To produce a large variety of cases, %we conduct two series of tests. 
% The first category corresponds to the computation of $\sigma_{r}$ and $\delta r$ for both the ground and space-borne exprimental set-ups, for multiple values of $r_{\rm true}$, and assuming the same correction scheme as in \cite{Morshed:2024fow}, i.e.:
% \begin{equation}
%     \left( \mathbf{\tilde{C}} \right)_{\ell} = C_{\ell}^{BB, \rm lens}.
% \end{equation}
% These are compared with the bias and standard deviation obtained from the sampling of the averaged likelihood \ref{eq:MI likelihood avg}. The results for these cases are described in FIG.~\ref{fig:validation vs rtrue}. We see that the statistical error increases with $r_{\rm true}$ which is due to the cosmic variance that grows with the amplitude of the power-spectrum. Similarly, the bias increases with $r_{\rm true}$ as expected, and the performance of the method to properly measure $r$ can be understood as a competition between the scaling of $\sigma_{r}$ and $\delta r$. Overall, we see a very good agreement between the Fisher approach and the full sampling of the averaged likelihood.
%In the second class of tests,
we introduce a degree of freedom in the correction parameter of the averaged likelihood by varying the amplitude of the CMB $B$-mode covariance estimate,
\begin{equation}
    \left( \Ctilde^{BB} \left( \alpha \right) \right)_{\ell} = \alpha \cdot C_{\ell}^{BB, \rm lens}, \label{eq:Ctilde alpha}
\end{equation}
as it was shown in \cite{Leloup:2023vkb} that when $\alpha = 0$, the bias on $r$ is
%considerable
significant, while in the case $\alpha = 1$ it is negligible with respect to the statistical uncertainty. Results with various values of $\alpha$ are shown in FIG.~\ref{fig:comparison sampling vs forecast} in the context of the ground experiment set-up. This comparison shows that the simplified Fisher approach is a really good approximation of the averaged likelihood and is able to quantitatively recover not only the bias, but also the covariance of the
%free
varying parameters.

From these results, we conclude that the formalism developed in Section~\ref{section:General formalism} is a useful tool that can be used for our study of robustness of the method to the correction term, and that the noise and CMB averaged likelihood is approximately Gaussian near its peak for small and large biases which allow us to focus on the bias $\delta r$ and Fisher error $\sigma_{r}$.

\subsection{Exploration of the correction term's impact}

\subsubsection{CMB estimate amplitude}
\label{subsubsection:CMB estimate amplitude}

\begin{figure}[!htb]
\begin{center}
\includegraphics[width=\columnwidth]{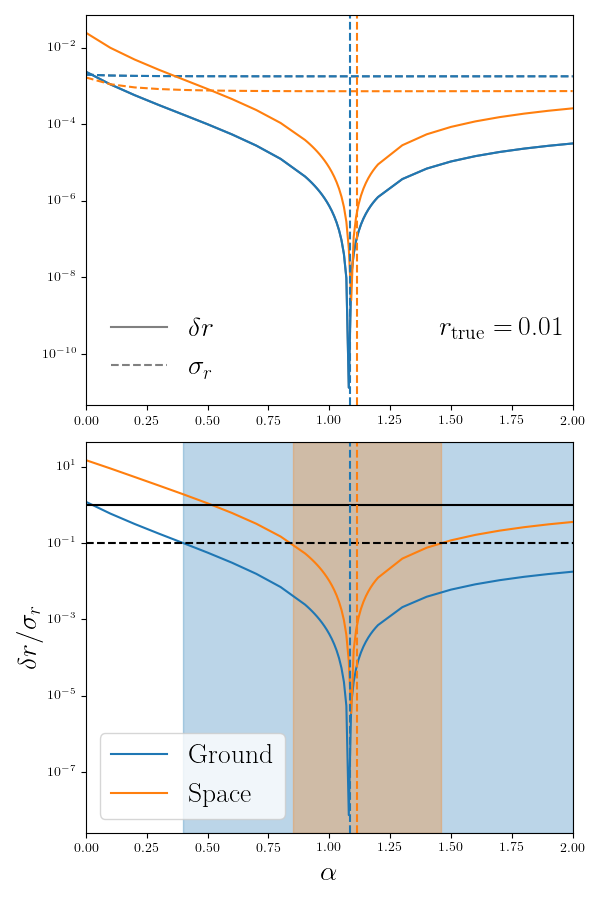}
\caption{Forecast of the MI approach robustness to a varying amplitude $\alpha$ of the correction term defined in Eq.~\eqref{eq:Ctilde alpha}, in terms of $\delta r$, $\sigma_{r}$ (Top) and their ratio (Bottom) calculated using the simplified Fisher forecast approximation described in Section~\ref{section:General formalism} for both the ground-based and space-borne instrumental configurations. The shaded areas correspond to the interval of $\alpha$ for which $\delta r \leq 0.1 \sigma_{r}$, and the vertical dashed lines to the estimation of $\alpha_{\rm min}$ from Eq.~\eqref{eq:alpha min estimation}. The solid and dashed horizontal lines correspond to $\delta r/\sigma_{r} = 1$ and 0.1 respectively. In this figure, the true value of the tensor-to-scalar ratio is assumed to be $r_{\rm true} = 0.01$. Results exhibit the same qualitative behavior for the other values tested, $r_{\rm true} = 0.0 \text{ and } 0.001$.}
\label{fig:perf vs alpha}
\end{center}
\end{figure}

Based on the implementation presented in Section~\ref{subsection:Implementation of the formalism}, we can further study how the performance of the MI method behaves, in terms of $\delta r$ and $\sigma_{r}$, as the CMB estimate $\Ctilde$ entering the correction term changes. We start by studying the performance of the method in both the ground based and space-borne experimental set-ups while varying the amplitude of $\Ctilde$, following the parameterization given in Eq.~\eqref{eq:Ctilde alpha}. Results are found in FIG.~\ref{fig:perf vs alpha} for a value of $r_{\rm true}=0.01$.

We see that for the bias to be small as compared to the standard deviation, using the 10\% threshold as an indicator, the correction needs not be too close to the lensing power-spectrum, especially for our ground based experimental set-up. The satellite experiment can tolerate a range of $\alpha \in \left[ 0.85, 1.46 \right]$, so with a width of $\sim 0.6$, while the ground experiment can withstand a range of $\alpha$ from $\sim 0.4$ all the way to the upper limit of our interval of exploration  $\alpha = 2$.
%$\sim 0.5$, for $r_{\rm true} = 0.01$.
The wider range for the latter compared to the former is driven by the fact that the bias is larger in general, and to a lesser extent because the sensitivity is better for the space-borne set-up. This is due to the larger $\ell$ range accessible from space, given that we use the same $\ell_{\rm max}$, and to the higher-dimensionality of the parameter space since there are many more frequency channels. For $r_{\rm true} = 0.001$ or $0$, the range is similar in width for both the ground and the space-borne experiment.
%, however the bias vanishes for the ground experiment so all tested values of $\alpha$ can be used.

An interesting fact that we can see here is that the actual minimum of the bias is not for $\alpha = 1$ where the correction term is obtained by using the exact lensing power spectrum, but for some $\alpha_{\rm min} > 1$. This can be understood by looking at the value of the gradient of the averaged likelihood at the true value of the parameters, $\mathbf{B} = \mathbf{\Hat{B}}$ and $\mathbf{C} = \mathbf{\Hat{C}}$,
\begin{eqnarray}
    && \left. \left< \frac{\partial \mathcal{S}_{\rm MI}}{\partial \beta} \right> \right|_{\beta, r = \beta_{\rm true}, r_{\rm true}} = \mathrm{Tr} \left[ \left( \mathbf{C} + \Nc \right)^{-1} \Nc_{, \beta} \right. \nonumber \\
    && \qquad \qquad \left. - \left( \mathbf{\tilde{C}} + \Nc \right)^{-1} \Nc_{, \Hat{\beta}} \right] \nonumber \\
    && = \sum_{\ell, m} \mathrm{Tr} \left[ \left( \left( r_{\rm true} \mathbf{C}_{\ell}^{\rm prim} + \mathbf{C}_{\ell}^{\rm lens} + \Nc \right)^{-1} \right. \right. \nonumber \\
    && \qquad \qquad \left. \left. - \left( \alpha \mathbf{C}_{\ell}^{\rm lens} + \Nc \right)^{-1} \right) \Nc_{, \beta} \right], \label{eq:bias} \\ [0.2cm]
    && \left. \left< \frac{\partial \mathcal{S}_{\rm MI}}{\partial r} \right> \right|_{\beta, r = \beta_{\rm true}, r_{\rm true}} = 0.
\end{eqnarray}

Therefore there is an $\alpha_{\rm min} \left( \beta_{\rm true}, r_{\rm true} \right)$ that is $>1$ for positive values of $r_{\rm true}$, that exactly cancels the right hand side of Eq.~\eqref{eq:bias}. In general, this minimum value will be different for each spectral parameter $\beta$, except when $r_{\rm true} = 0$ for  which there is a common trivial solution $\alpha_{\rm min} = 1$. So, for a given $r_{\rm true}$ the optimal value will be a compromise $\alpha_{\rm min} \geq 1$ that does not completely cancel the bias. However, because $\Nc_{, \beta}$ is independent of $\ell$ in our case, $\alpha_{\rm min}$ is the same for each spectral parameter, and the bias is exactly canceled. Of course, because it depends on the true values of the parameters, this optimal value of $\alpha$ that minimizes the bias is not known in advance. A good and intuitive approximation to find $\alpha_{\rm min}$ in the case under study, whose result is included in FIG.~\ref{fig:perf vs alpha}, is to assume that $r_{\rm true} \mathbf{C}_{\ell}^{\rm prim} + \mathbf{C}_{\ell}^{\rm lens} + \Nc$ is constant with respect to $\ell$ which is approximately the case past the reionization bump and until the first peak of the lensing spectrum. Under this approximation, the minimum is found at
\begin{equation}
    \alpha_{\rm min} \simeq 1 + r_{\rm true} \frac{\displaystyle \sum_{\ell} \left( 2\ell + 1 \right) \mathbf{C}_{\ell}^{\rm prim}}{\displaystyle \sum_{\ell} \left( 2\ell + 1 \right) \mathbf{C}_{\ell}^{\rm lens}}, \label{eq:alpha min estimation}
\end{equation}
where the sum runs over the multipole range accessible to the experiment. This assumption is best fulfilled in the range of multipoles where $\mathbf{C}_{\ell}^{\rm lens} + \Nc$ dominates, which makes the above approximation better as the true value of the tensor-to-scalar ratio $r_{\rm true}$ is small. It is a particularly good approximation for the Ground experiment even at higher values of $r_{\rm true}$ as the largest scales, at which the primordial contribution dominates, are not included in its accessible multipole range.

\subsubsection{Multipole range}
\label{subsubsection:Multipole range}

Another degree of freedom of the correction term one can play with is the multipole range over which it is applied, in particular the lowest multipole $\ell_{\rm min}$ from which the CMB estimate is considered, i.e. such that
\begin{equation}
    \left( \Ctilde^{BB} \left( \ell_{\rm min} \right) \right)_{\ell} = \begin{cases}
      0 & \text{if $\ell \leq \ell_{\rm min}$}\\
      C_{\ell}^{BB, \rm lens} & \text{otherwise}
    \end{cases} \label{eq:Ctilde lmin}
\end{equation}

The decision to study this parameter is motivated by the fact that the primordial part of the $B$-mode power-spectrum is dominant at low multipoles, while it is small compared to the lensing power-spectrum at higher multipoles. Therefore, by canceling the correction term at multipoles below $\ell_{\rm min}$, the range $\ell \leq \ell_{\rm min}$ can be understood as the signal region in which the tensor-to-scalar ratio is measured. On the other hand, the range $\ell \geq \ell_{\rm min}$ is interpreted as a side-band where the foreground parameters are measured. Although this is an intuitive interpretation, it is also too simplistic as the side-band also brings some contribution to the constraint on $r$ and the signal region helps measuring foreground parameters.

\begin{figure}[!htb]
\begin{center}
\includegraphics[width=\columnwidth]{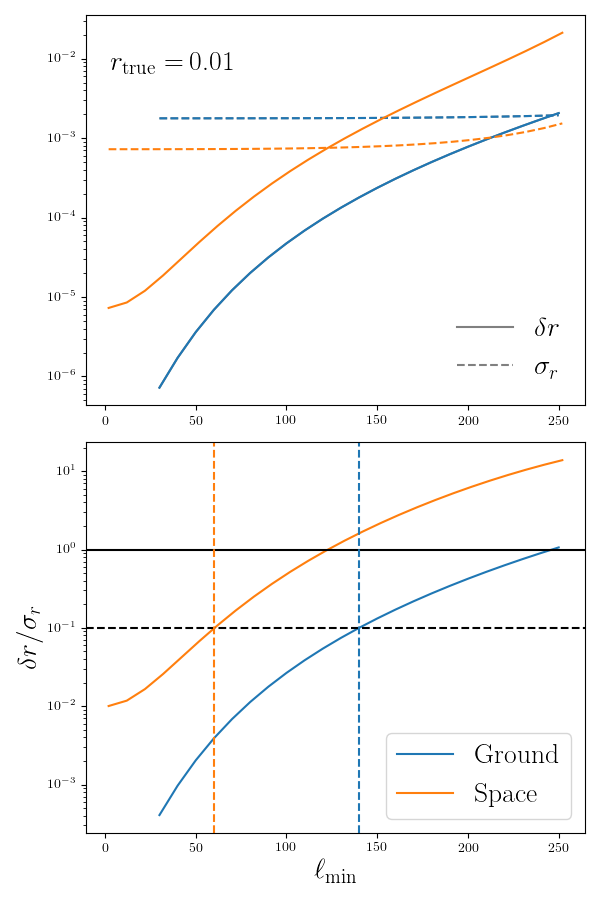}
\caption{Forecast of the MI approach robustness to a varying lowest multipole $\ell_{\rm min}$ of the correction term defined in Eq.~\eqref{eq:Ctilde lmin}, in terms of $\delta r$, $\sigma_{r}$ (Top) and their ratio (Bottom) calculated using the simplified Fisher forecast approximation described in Section~\ref{section:General formalism} for both the ground-based and space-borne instrumental configurations. The vertical dashed lines correspond to the maximum multipole $\ell_{\rm min}$ such that $\delta r \leq 0.1 \sigma_{r}$, which are at $\ell_{\rm min} = 134$ and $\ell_{\rm min} = 71$ for ground and space experiments respectively. The solid and dashed horizontal lines correspond to $\delta r/\sigma_{r} = 1$ and 0.1 respectively. In this figure, the true value of the tensor-to-scalar ratio is assumed to be $r_{\rm true} = 0.01$.  Results exhibit the same qualitative behavior for the other values tested, $r_{\rm true} = 0.0 \text{ and } 0.001$.}
\label{fig:perf vs lmin}
\end{center}
\end{figure}

The performance of the method in terms of $\delta r$ and $\sigma_{r}$ as a function of $\ell_{\rm min}$ is shown in FIG.~\ref{fig:perf vs lmin} for both experimental set-up and $r_{\rm true} = 0.01$. It appears that there is a range of $\ell_{\rm min} > 0$ such that the bias is $\delta r \leq 0.1 \sigma_{r}$ when $r_{\rm true} = 0.01$, the highest such multipole being 140 and 60 for ground and space experiments respectively. This limit multipole is rather stable as
the true value of the tensor-to-scalar ratio decreases, with limits at 147 and 66 when $r_{\rm true} = 0.001$, and 148 and 64 when $r_{\rm true} = 0$.

\begin{figure*}[!htb]
\begin{center}
\includegraphics[width=1.7\columnwidth]{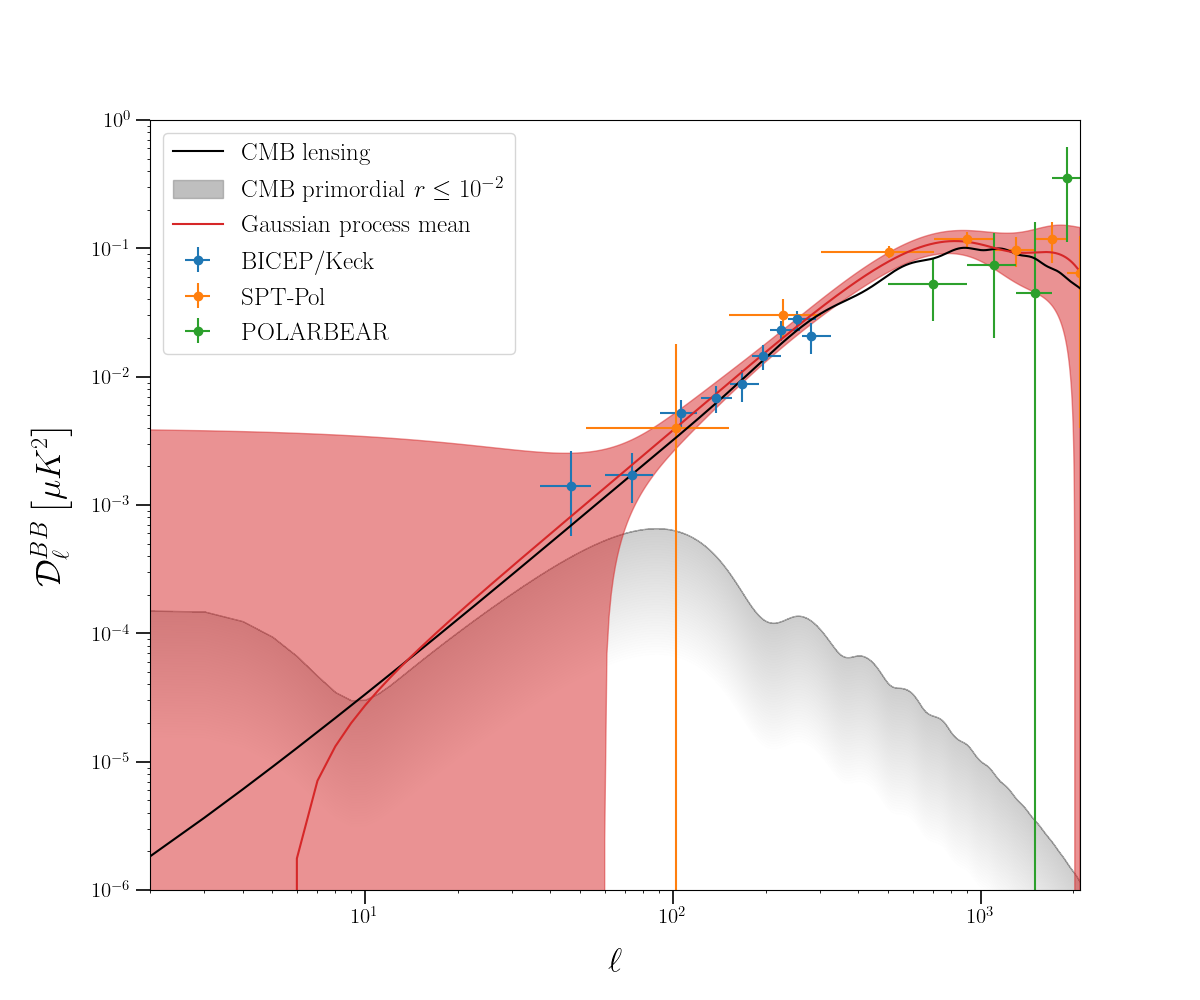}
\caption{Theoretical $B$-mode power spectra of the CMB lensing and primordial gravitational waves for values of $r \leq 10^{-2}$ used to produce the noiseless data covariance $\mathbf{D}$ and used as reference for the CMB estimate $\Ctilde$ of the correction term. The data points are current best measurements from BICEP/Keck~\cite{BICEP:2021xfz}, SPT-Pol~\cite{SPT:2019nip} and POLARBEAR~\cite{POLARBEAR:2017beh}, with their error bars, used to produce the Gaussian process mean estimate (red solid line) and its 1$\sigma$ confidence interval (shaded red area).}
\label{fig:DlBB data}
\end{center}
\end{figure*}

We see that the performance of the method decreases as $\ell_{\rm min}$ increases which is easily understood since as $\ell_{\rm min}$ grows, the contribution of more and more multipoles is not corrected.

\subsubsection{From available data}
\label{subsubsection:From available data}

From the previous results obtained in this section, it appears that there is some tolerance to the correction term in order to make the inherent bias from the method small as compared to the statistical error. A legitimate approach to try, based on this general result, is whether the $B$-mode data currently available is enough to produce a model-independent estimate of the CMB covariance $\Ctilde$ that can be used in the correction term and can guarantee that $\delta r$ will be small enough that it is not a problem to be worried about.

To explore this approach, we use the measured binned $B$-mode power-spectra from three datasets: BICEP/Keck 2018 observation season \cite{BICEP:2021xfz}, SPT-Pol \cite{SPT:2019nip} and the POLARBEAR 2-year data release \cite{POLARBEAR:2017beh}. These give tight constraints on the shape of the $B$-mode CMB power-spectrum, \textit{a priori} coming from the gravitational lensing of $E$ modes, over the wide range of multipoles $37 \leq \ell \leq 2302$. This interval far extends the multipole range used in the forecast, however we noticed that the inclusion of each of the three datasets has a noticeable impact, even at low multipoles.

Because we want this estimate to be model-independent, we need a tool to produce a function of $\ell$ based solely on data points. We therefore choose to use Gaussian Processes as a simple way to achieve this goal. We use the Gaussian Process package of \texttt{scikit-learn} \cite{scikit-learn} with the built-in radial basis function kernel operator to produce the data-based $\Ctilde$ since it can handle data points with error-bars. However, its implementation requires the error-bars to be symmetrical, therefore we simply decided to symmetrize them by taking the maximum of the lower and upper errors from the data. We also neglect the correlations between the data points of single experiments, and correlations between experiments. The CMB power-spectrum estimated from existing datasets is shown in FIG.~\ref{fig:DlBB data} along with the data points and the theoretical spectra used to produce the CMB part of the noiseless data variance $\mathbf{D}$.

\begin{figure}[!htb]
\begin{center}
\includegraphics[width=\columnwidth]{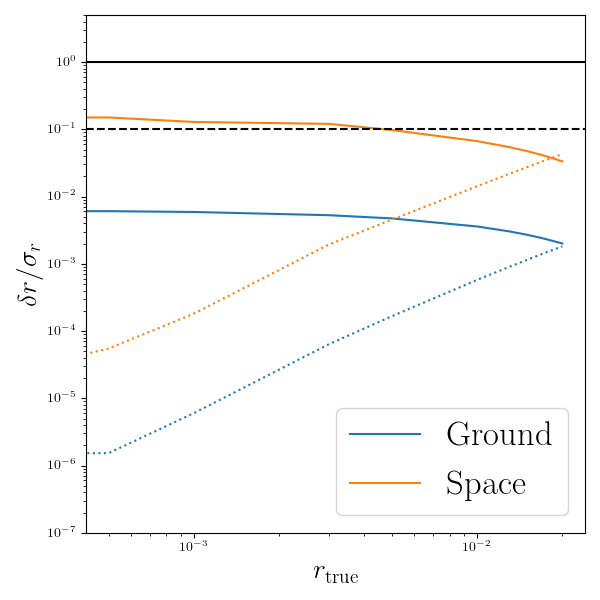}
\caption{Forecast of the MI approach robustness as a function of $r_{\rm true}$, in terms of $\delta r$, $\sigma_{r}$ (Top) and their ratio (Bottom) calculated using the simplified Fisher forecast approximation described in Section~\ref{section:General formalism} for both the ground-based and space-borne instrumental configurations. The CMB estimates $\Ctilde$ used in the correction term are the mean of the Gaussian process produced from the datasets described in Section~\ref{subsubsection:From available data} (solid) and the CMB lensing power-spectrum itself (dotted).
%and the re-scaled lensing spectrum with $\alpha = \alpha_{\rm min}$ determined using Eq.~\eqref{eq:alpha min estimation} (dashed).
The solid and dashed horizontal lines correspond to $\delta r/\sigma_{r} = 1$ and 0.1 respectively.}
\label{fig:deltar data}
\end{center}
\end{figure}

We use the developed forecast framework to study the performance of the MI approach when using the mean of the Gaussian process result in place of our estimate of the CMB $B$-mode covariance, as a function of the true value of the tensor-to-scalar ratio. As a mean of comparison, we run through the same cases using $\left( \Ctilde^{BB} \right)_{\ell} = C_{\ell}^{BB, \rm lens}$.
%, and $\left( \Ctilde \right)_{\ell} = \alpha_{min} \cdot C_{\ell}^{BB, \rm lens}$ with $\alpha_{\rm min}$ determined by the approximation of Eq.~\eqref{eq:alpha min estimation}. Although this approximation requires to know $r_{\rm true}$, the interpretation of this estimation scheme is straightforward and intuitive, and it has been demonstrated in Section~\ref{subsubsection:CMB estimate amplitude} to display very good performance.
The results of this comparative evaluation can be found in FIG.~\ref{fig:deltar data}.

%As expected from the results of previous sections, the case where the correction term is taken from the re-scaled lensing power-spectrum with $\alpha = \alpha_{\rm min}$ performs better than the case using the Gaussian process over almost the whole range of $r_{\rm true}$, the performance being worse as $r_{\rm true}$ increases since the approximation for $\alpha_{\rm min}$ gets less accurate. However, it
It is reassuring to see that the performance using the Gaussian process is good in general, with $\delta r < 0.1 \sigma_{r}$ over the whole range of $r_{\rm true}$ for the ground experiment and $\delta r < 0.2 \sigma_{r}$ for the space mission,
%while it is $\delta r < 0.2 \sigma_{r}$ for the satellite,
the largest biases being at low values of $r_{\rm true}$.

The bias is driven by the fact that the $B$-mode power-spectrum of the Gaussian process mean has a larger amplitude than the CMB lensing spectrum and vanishes for the lowest multipoles, i.e. a combination of both effects studied previously corresponding to an effective $\alpha$ slightly larger than 1 and an effective $\ell_{\rm min} \simeq 5-10$. The latter effect has no impact on the ground based experimental set-up as its multipole range includes only $30 \leq \ell$, while we have seen in FIG.~\ref{fig:perf vs alpha} that the former effect has a bigger impact on the space experiment.

\section{Validation and discussions}
\label{section:Validation and discussions}

We have seen in the previous Section how the forecasting simplified approach can be used to study the dependency of the performance as the CMB estimate $\Ctilde$ that enters in the regularization of the likelihood changes. We have found that, taking as criterion $\delta r \leq 0.1 \sigma_{r}$, the method can accommodate for a range of amplitude mismatch with the true power-spectrum. The performance is also good when the regularization scheme is not applied to the whole range of multipoles. From these results, we could understand intuitively how using a Gaussian process produced from current observations of CMB $B$ modes as $\Ctilde$ seemed to be already enough to lead to low bias, for both the ground experiment and space-borne mission.

However, from the limitations of the forecasting approach mentioned at the end of Section~\ref{section:General formalism}, it is necessary to validate these results using an established implementation likely to be used in actual data analysis, following the pipeline described in FIG.~\ref{fig:analysis pipeline}. To this end, we explore a small number of configurations corresponding to cases described in Section~\ref{section:Dependence on the correction term} on a few simulations of CMB and noise, with the \texttt{MICMAC}~\cite{Morshed:2024fow} package.

\subsection{Choice of the case study}
\label{subsection:Choice of the case study}

In order to validate the forecasting approach with \texttt{MICMAC} (right side of the pipeline in FIG.~\ref{fig:analysis pipeline}), a total of six different configurations for $\Ctilde$ are considered with a space mission configuration and input tensor-to-scalar ratio value taken to be $r_{\rm true} = 0.01$. To avoid potential issues with the masking procedure, we perform the component separation on the full sky. Given how we account for the sky fraction in the forecasting approach, we expect that this will not affect the bias and will reduce the statistical error by a factor of $\sqrt{f_{\rm sky}} \sim 0.7$ as compared to a space mission observing half the sky.

%~\cite{SimonsObservatory:2018koc}for their observed frequency channels, noise levels and are run with mask of coverage $f_{\rm sky} \sim 10 \%$. 
All corresponding runs are performed on HEALPix~\cite{Healpix} grid maps featuring a resolution of $\mathrm{nside}=128$ and a maximum multipole of $\ell_{\rm max}=256$. 

As the $\texttt{MICMAC}$ package is computationally expensive, a quantitative validation based on the averaging over many different realizations and Gibbs chains is not feasible. The validation methodology adopted is instead based on the repeatability of the sampling steps given different correction procedures, with a more qualitative validation of the forecasting approach.  

For each configuration, five sets of input maps are generated as linear mixture of random CMB realizations from the theoretical power spectra, Gaussian noise obtained from the space mission noise levels, and \texttt{d0s0} foreground maps as provided by the \texttt{PySM} software. To each realization is associated a fixed random seed to perform the Gibbs sampling, ensuring that each correction configuration runs with the exact same of parameters over the exact same five set of simulated input frequency maps, differing only by the choice of the CMB estimate $\Ctilde$.

As previously noted, the input simulated maps are computed as a set of $Q$ and $U$ pixel maps, and $E$ and $B$ modes are accounted for in the sampling. We follow the same procedure as in the previous sections, fitting for cosmological parameters and varying the correction term only for the $B$ modes.
%In particular, the previous case studied in Section~\ref{section:Dependence on the correction term} were done with variation on $\Ctilde$ as a $B$-mode spectrum only, while a typical \texttt{MICMAC} run also requires an $E$-mode $\Ctilde$ spectrum. 

%In the following, the $\Ctilde^{\rm EE}$ spectrum is always taken to be the $E$-mode spectrum computed with \texttt{CAMB}, and which is also used to generate the input CMB maps. 

\begin{figure*}%[!htb]
\begin{center}
\includegraphics[width=0.8\textwidth]{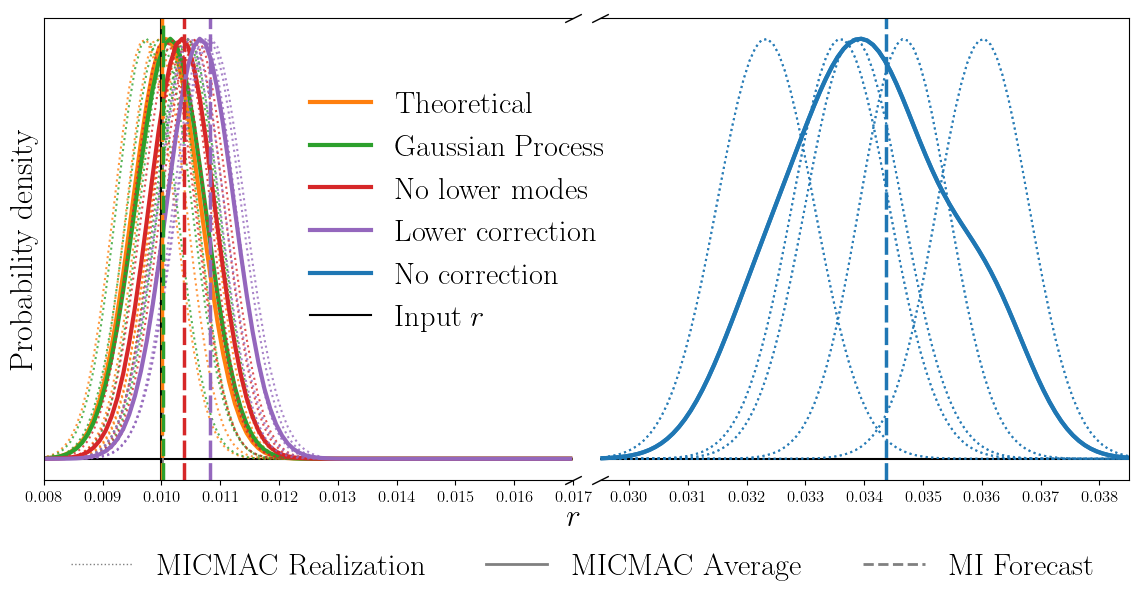}
\caption{Probability densities of the $r$ posterior for each configuration described in Section~\ref{subsection:Choice of the case study}. For each configuration, the solid bold line is the average of the probability densities represented as thin solid lines of the same color. For clarity, the $x$-axis is cut between $r=0.017$ and $r=0.0295$. For each configuration, the vertical dashed lines denote the corresponding forecast bias predictions. The \textbf{Theoretical} and Gaussian Process cases bias predictions are very small compared to the statistical error.
%Probability densities of the $r$ posterior for each configuration described in Section~\ref{subsection:Choice of the case study}, with the \texttt{Theoretical} involving a theoretical input CMB spectrum being chosen as $\Ctilde$, \texttt{Data-driven} the configuration 5. involving data-based Gaussian process configuration, \texttt{$\ell_{\rm min}=100$} the case 4. with $\Ctilde^{\rm BB}_{\ell < 100} = 0$, and the two last $\Ctilde$ configurations 5. and 1. being parametrized by Eq.~\ref{eq:Ctilde alpha} and denoted by the value of \texttt{$\alpha$}. 
}
%https://www.canva.com/design/DAGShBWHkdE/Yu4joJVEiLO_w_vMdKSwXw/edit?utm_content=DAGShBWHkdE&utm_campaign=designshare&utm_medium=link2&utm_source=sharebutton
\label{fig:micmac validation}
\end{center}
\end{figure*}

The different choices of configurations for
%only differ from the choice of
$\Ctilde^{BB}$ are:
%, taken to be either:
\begin{enumerate}
    \item \textbf{No correction:} without correction, i.e. taking $\alpha= 0$ in Eq.~(\ref{eq:Ctilde alpha}), as a reference biased case, with a bias predicted by the forecasting approach to be of the order $\delta_r \sim 10 \sigma_r \sim 3 \times 10^{-2}$. As the implementation in MICMAC can not handle the exact case $\alpha=0$ for numerical reasons, we instead take $\alpha=10^{-10}$ which should exhibit the same behavior, 
    \item \textbf{Theoretical:} taking the theoretical $B$-mode lensing power spectrum computed by \texttt{CAMB}, which is the $\Ctilde^{BB}$ that would be taken for a typical \texttt{MICMAC} run, associated to a prediction of $\delta_r \sim 10^{-2} \sigma_r \sim 10^{-5}$,
    \item \textbf{Lower correction:} taking the $BB$ lensing power spectrum computed by \texttt{CAMB} multiplied to $\alpha= 0.5$ as in Eq.~(\ref{eq:Ctilde alpha}) and studied with the forecasting tool in Section~\ref{subsubsection:CMB estimate amplitude}, with a predicted bias $\delta_r \sim \sigma_r / 2 \sim 5 \times 10^{-4}$,
    \item \textbf{No lower modes:} taking the $BB$ lensing power spectrum computed by \texttt{CAMB} with $\ell_{\rm min} = 100$, as in Section~\ref{subsubsection:Multipole range}, predicted to be biased as $\delta_r \sim \sigma_r / 5 \sim 2 \times 10^{-4}$,
    \item \textbf{Gaussian Process:} taking the data-based Gaussian process power spectrum studied in Section~\ref{subsubsection:From available data}, with $\delta_r \sim 5\times10^{-3} \sigma_r \sim 5 \times 10^{-6}$.
\end{enumerate}
%considering the statistical error given in Figure~\ref{fig:perf vs alpha} as $\sigma_r = 10^{-3}$ for $r_{\rm true} = 0.01$.
As in the previous sections, the fixed estimate of the CMB $E$ modes is taken to be identical as the input CMB power spectrum. 

% Talk about what we expect maybe?

It should be emphasized that the validation procedure relies here on a limited number of realizations, and as such will not be able to recover precisely the lowest $\delta_r$ biases compared to $\sigma_r$. In particular, we expect to retrieve the same order of magnitude and not the exact value of $\delta_r$ mentioned above.

\subsection{Results and discussions}

The results are summarized in FIG.~\ref{fig:micmac validation}, displaying the different $r$ posteriors for each of the configuration.

The theoretical $\Ctilde$ and the Gaussian process cases are consistent with the input $r$ value.
Both configurations have very close posteriors, which is expected as their corresponding biases are negligible compared to the statistical error on $r$ and all configurations depart from the same seed. 

The $\ell_{\rm min} = 100$ configuration is also compatible with the input $r$ value, although slightly biased with the five realizations giving $\delta_r^{\, \rm{\ell_{min}}=100} = 4 \times 10^{-4}$. This bias is consistent with the expectation of $\delta_r \sim \sigma_r / 5$. 

The configurations parametrized by $\alpha$ as in Eq.~(\ref{eq:Ctilde alpha}) are associated to a greater bias, with $\delta_r^{\, \alpha = 0.5} = 6 \times 10^{-4}$ and $\delta_r^{\, \alpha = 10^{-10}} = 2.5 \times 10^{-2}$, which is consistent with the predictions of $\delta_r \sim \sigma_r / 2$ and $\delta_r \sim 10 \sigma_r$ respectively. 

As the three cases $\ell_{\rm min}=100$, $\alpha = 0.5$ and $\alpha = 10^{-10}$ featuring large biases correspond in fact to quite poor choices of $\Ctilde^{BB}$, these results highlight a relatively important margin in the choice of $\Ctilde^{BB}$, and the relevance of a data-driven choice. 
We mention as well that the ground based experiment is predicted to be much less impacted by a poor choice of $\Ctilde^{BB}$ since the smaller observed sky fraction means that less modes contribute to the bias and the statistical error is increased.

The study of the possible deviation of fixed CMB estimate for the $E$ modes is left for future studies focused on the retrieval of $E$ modes, both in harmonic and pixel domain. 
While we expect a deviation of $\Ctilde^{EE}$ to
%possibly highly
impact the retrieved parameters as more signal is not canceled out, we also expect to have a better signal over noise ratio to better characterize the $E$-mode power spectrum. 
% Last sentence maybe not necessary

\section{Conclusions}
\label{section:Conclusions}

In the present work, we developed and validated a forecasting approach for the minimally informed component separation method based on the Fisher formalism. This approach is able to quantitatively estimate the bias and statistical error of cosmological parameters for very little computational cost. Although currently limited to foreground models with constant SEDs across the sky, it is expected that this forecasting tool will render possible future impact studies of instrumental systematic effects on the performance of the MI method, as well as to optimize its hyper-parameters and validate its general assumptions.

A general analysis pipeline for such studies has been proposed, capitalizing on both the fast estimation by the forecasting tool and the accurate but more involved evaluation by the full pixel-based package \texttt{MICMAC}. As an illustration, we followed this procedure to study the impact of the CMB covariance estimate $\Ctilde$ on the performance of the MI method, and in particular as a regularization for its bias. We were able to find a wide variety of configurations for $\Ctilde$ departing from the theoretical CMB lensing $B$-mode covariance and such that the bias is unnoticeable given the statistical error. This builds confidence that the accuracy of $\Ctilde$ necessary for the ad-hoc bias correction procedure will not be a limiting factor when applied to actual observations, strengthening the case for the MI approach.

\section*{Acknowledgements}

%We would like to thank Davide Poletti for his constribution to early discussions about this project, and Dominic Beck for reading the manuscript. The authors acknowledge support of the French National Research Agency (Agence National de la Recherche) grants: ANR BxB and ANR B3DCMB. This work is also part of the SCIPOL project funded by the European Research Council (ERC) under the European Union’s Horizon 2020 research and innovation program (PI: Josquin Errard, Grant agreement No. 101044073). Some of the results in this paper have been derived using the \texttt{healpy} \cite{2005ApJ...622..759G,Zonca2019,healpix}, \texttt{numpy} and \texttt{PySM} packages. Some of the figures in this article have been created using \texttt{GetDist}.

% Former acknowledgements for MICMAC: 
We would like to thank Radek Stompor and Josquin Errard for fruitful discussions and feedback on the draft. The authors acknowledge the use of the HEALPix~\cite{Healpix} and \texttt{healpy} packages \cite{Zonca2019,Healpy}, and the \texttt{JAX} package~\cite{jax2018github}. 

% For JZ: 
This work was granted access to the HPC resources of IDRIS under the allocation 2024-102865 made by GENCI. These resources were allocated to the SCIPOL project~\cite{Scipol} funded by the European Research Council (ERC) under the European Union’s Horizon 2020 research and innovation program (PI: Josquin Errard, Grant agreement No. 101044073). This work has received additional funding by the European Union’s Horizon 2020 research and innovation program under grant agreement no. 101007633 CMB-Inflate. This work was supported by JSPS KAKENHI Grant Number 25K17407.

% For Martina's ERC: 
MM is funded by the European Union (ERC, RELiCS, project number 101116027). Views and opinions expressed are however those of the author(s) only and do not necessarily reflect those of the European Union or the European Research Council Executive Agency. Neither the European Union nor the granting authority can be held responsible for them.

%\section*{Bibliography}
%\vspace{-0.8cm}
%\bibliographystyle{ieeetr}
\bibliography{bibliography}

%\clearpage
\appendix

\section{Derivatives entering the Fisher information matrix}
\label{appendix:Derivatives entering the Fisher information matrix}

We describe here the first and second derivatives of the noise after component separation $\Nc$ and of the difference between the estimated and true CMB covariance $\hatD$ with respect to the foreground parameters $\beta$ that enter the expression of the Fisher information matrix elements in Eq.~\eqref{eq:d2S/db2}-\eqref{eq:d2S/dgamma2}:
\begin{widetext}
\begin{eqnarray}
    && \Nc = \mathbf{E}^{T} \left( \mathbf{B}^{T} \mathbf{N}^{-1} \mathbf{B} \right)^{-1} \mathbf{E}, \\
    && \Nc_{,\beta} = - \mathbf{E}^{T} \left( \mathbf{B}^{T} \mathbf{N}^{-1} \mathbf{B} \right)^{-1} \left( \mathbf{B}_{,\beta}^{T} \mathbf{N}^{-1} \mathbf{B} + \mathbf{B}^{T} \mathbf{N}^{-1} \mathbf{B}_{,\beta} \right) \left( \mathbf{B}^{T} \mathbf{N}^{-1} \mathbf{B} \right)^{-1} \mathbf{E}, \\
    && \Nc_{,\beta\beta'} = 2\, \mathbf{E}^{T} \left( \mathbf{B}^{T}\mathbf{N}^{-1}\mathbf{B} \right)^{-1} \left[ - \mathbf{B}_{,\beta\beta'}^{T}\mathbf{N}^{-1}\mathbf{B} + \mathbf{B}^{T}\mathbf{N}^{-1}\mathbf{B}_{,\beta} \left( \mathbf{B}^{T}\mathbf{N}^{-1}\mathbf{B} \right)^{-1} \mathbf{B}_{,\beta'}^{T}\mathbf{N}^{-1}\mathbf{B} - \mathbf{B}_{,\beta}^{T}\mathbf{P}\mathbf{B}_{,\beta'} \right. \nonumber \\
    && \left. \qquad +\, \mathbf{B}^{T}\mathbf{N}^{-1}\mathbf{B}_{,\beta'} \left( \mathbf{B}^{T}\mathbf{N}^{-1}\mathbf{B} \right)^{-1} \mathbf{B}^{T}\mathbf{N}^{-1}\mathbf{B}_{,\beta} + \mathbf{B}^{T}\mathbf{N}^{-1}\mathbf{B}_{,\beta} \left( \mathbf{B}^{T}\mathbf{N}^{-1}\mathbf{B} \right)^{-1} \mathbf{B}^{T}\mathbf{N}^{-1}\mathbf{B}_{,\beta'} \right] \left( \mathbf{B}^{T}\mathbf{N}^{-1}\mathbf{B} \right)^{-1} \mathbf{E}. \\[10pt]
    && \hatD = \mathbf{C} - \mathbf{E}^{T} \left( \mathbf{B}^{T} \mathbf{N}^{-1} \mathbf{B} \right)^{-1} \mathbf{B}^{T} \mathbf{N}^{-1} \mathbf{D} \mathbf{N}^{-1} \mathbf{B} \left( \mathbf{B}^{T} \mathbf{N}^{-1} \mathbf{B} \right)^{-1} \mathbf{E}, \\
    && \hatD_{, \beta} = 2\, \mathbf{E} \left( \mathbf{B}^{T} \mathbf{N}^{-1} \mathbf{B} \right)^{-1} \left( \mathbf{B}^{T} \mathbf{N}^{-1} \mathbf{B}_{,\beta} \left( \mathbf{B}^{T} \mathbf{N}^{-1} \mathbf{B} \right)^{-1} \mathbf{B}^{T} \mathbf{N}^{-1} - \mathbf{B}_{, \beta}^{T} \mathbf{P} \right) \mathbf{D} \mathbf{N}^{-1} \mathbf{B} \left( \mathbf{B}^{T} \mathbf{N}^{-1} \mathbf{B} \right)^{-1} \mathbf{E}, \\
    && \hatD_{, \beta\beta'} = 2\, \mathbf{E} \left( \mathbf{B}^{T} \mathbf{N}^{-1} \mathbf{B} \right)^{-1} \left[ \left( \mathbf{B}^{T} \mathbf{N}^{-1} \mathbf{B}_{, \beta\beta'} \left( \mathbf{B}^{T} \mathbf{N}^{-1} \mathbf{B} \right)^{-1} \mathbf{B}^{T} \mathbf{N}^{-1} - \mathbf{B}_{, \beta\beta'}^{T} \mathbf{P} + \mathbf{B}_{, \beta}^{T} \mathbf{P} \mathbf{B}_{, \beta'} \left( \mathbf{B}^{T} \mathbf{N}^{-1} \mathbf{B} \right)^{-1} \mathbf{B}^{T} \mathbf{N}^{-1} \right. \right. \nonumber \\
    && + \mathbf{B}_{, \beta'}^{T} \mathbf{P} \mathbf{B}_{, \beta} \left( \mathbf{B}^{T} \mathbf{N}^{-1} \mathbf{B} \right)^{-1} \mathbf{B}^{T} \mathbf{N}^{-1} + \left( \mathbf{B}_{, \beta}^{T} \mathbf{N}^{-1} \mathbf{B} + \mathbf{B}^{T} \mathbf{N}^{-1} \mathbf{B}_{, \beta} \right) \left( \mathbf{B}^{T} \mathbf{N}^{-1} \mathbf{B} \right)^{-1} \mathbf{B}_{, \beta'}^{T} \mathbf{P} \nonumber \\
    && + \left( \mathbf{B}_{, \beta'}^{T} \mathbf{N}^{-1} \mathbf{B} + \mathbf{B}^{T} \mathbf{N}^{-1} \mathbf{B}_{, \beta'} \right) \left( \mathbf{B}^{T} \mathbf{N}^{-1} \mathbf{B} \right)^{-1} \mathbf{B}_{, \beta}^{T} \mathbf{P} - \mathbf{B}_{, \beta}^{T} \mathbf{N}^{-1} \mathbf{B} \left( \mathbf{B}^{T} \mathbf{N}^{-1} \mathbf{B} \right)^{-1} \mathbf{B}^{T} \mathbf{N}^{-1} \mathbf{B}_{, \beta'} \left( \mathbf{B}^{T} \mathbf{N}^{-1} \mathbf{B} \right)^{-1} \mathbf{B}^{T} \mathbf{N}^{-1} \nonumber \\
    && \left. - \mathbf{B}_{, \beta'}^{T} \mathbf{N}^{-1} \mathbf{B} \left( \mathbf{B}^{T} \mathbf{N}^{-1} \mathbf{B} \right)^{-1} \mathbf{B}^{T} \mathbf{N}^{-1} \mathbf{B}_{, \beta} \left( \mathbf{B}^{T} \mathbf{N}^{-1} \mathbf{B} \right)^{-1} \mathbf{B}^{T} \mathbf{N}^{-1} \right) \mathbf{D} \mathbf{N}^{-1} \mathbf{B} \nonumber \\
    && \left. \left( \mathbf{B}^{T} \mathbf{N}^{-1} \mathbf{B}_{,\beta} \left( \mathbf{B}^{T} \mathbf{N}^{-1} \mathbf{B} \right)^{-1} \mathbf{B}^{T} \mathbf{N}^{-1} - \mathbf{B}_{, \beta}^{T} \mathbf{P} \right) \mathbf{D} \left( \mathbf{N}^{-1} \mathbf{B} \left( \mathbf{B}^{T} \mathbf{N}^{-1} \mathbf{B} \right)^{-1} \mathbf{B}_{, \beta'}^{T} \mathbf{N}^{-1} \mathbf{B} - \mathbf{P} \mathbf{B}_{, \beta'} \right) \right] \left( \mathbf{B}^{T} \mathbf{N}^{-1} \mathbf{B} \right)^{-1} \mathbf{E}.
\end{eqnarray}
\end{widetext}

\end{document}